\documentstyle[11pt]{article}
\begin{document}

\begin{titlepage}
\begin{flushright}
IFUG 95/11 \\
\today\\
\end{flushright}
\vspace{1 cm}
\begin{center}
\Large
{\bf Spin $3/2$ Fields Non-Minimal Coupling as Square Root}\\
{\bf of Linearized Gravity with Matter}\\
\vspace{.5 cm}
\normalsize
\large{J. A. Nieto}\footnote{e-mail: nie@zeus.ccu.umich.mx} \\
\baselineskip 24pt
\lineskip 10pt
\normalsize
\vspace{.5 cm}
{\em Escuela de Ciencias F\'{\i}sico-Matem\'aticas,\\
Universidad Michoacana, Apartado Postal 749, \\ 58000 Morelia Michoac\'an,
M\'exico }\\
\vspace{.5 cm}
\large{O. Obreg\'on \footnote{e-mail: octavio@ifug.ugto.mx} and
V. M. Villanueva \footnote{e-mail: victor@ifug2.ugto.mx}} \\
\baselineskip 24pt
\lineskip 10pt
\normalsize
\vspace{.5 cm}
{\em Instituto de F\'{\i}sica de la Universidad de Guanajuato, \\
Apartado Postal E-143, \\
37150 Le\'on Guanajuato, M\'exico. \\ and \\
Universidad Aut\'onoma Metropolitana-Iztapalapa,\\
Departamento de F\'{\i}sica, Apartado Postal 55-534 \\
D.F. M\'exico, M\'exico }\\
\vspace{.5 cm}
\end{center}
\begin{abstract}
\noindent
A non-minimal coupling for spin $3/2$ fields is
obtained. We use the fact that the Rarita-Schwinger field equations are the
square root of the full linearized Einstein field equations in order to
investigate the form of the interaction for the spin $3/2$ field with gauge
fields. We deduce the form of the interaction terms for the
electromagnetic and non-abelian Yang-Mills fields by implementing appropiate
energy momentum tensors on the linearized Einstein field equations.
The interaction found for the electromagnetic case happens to coincide with
the dipole term found by Ferrara {\it et al} by a very different procedure,
namely by demanding $g=2$ at the tree level for the electromagnetic
interaction of arbitrary spin particles.
The same kind of interaction is found by using the resource of linearized
Supergravity N=2.
For the case of the Yang-Mills field Supergravity N=4 is linearized, providing
also the already foreseen interaction.
\end{abstract}
\begin{center}
\vspace{0.5cm}
PACS ~~12.90.+b, 11.90.+t, 04.65.+e\\
\end{center}

\end{titlepage}

\baselineskip 24pt
\lineskip 10pt
\parskip=11pt

\def\theequation{\arabic{section}.\arabic{equation}}


\section{Introduction}
The standard treatment of the free masless spin $3/2$ field is achieved by
means of the Rarita-Schwinger (R-S) lagrangian \cite{RS,LL},
\begin{equation}
\label{1.1}
L_{RS} = - \frac {1}{2} \epsilon^{\mu\nu\rho\sigma}
\bar \Psi_\mu \gamma_5 \gamma_\nu \partial_\rho \Psi_\sigma \,,
\end{equation}
\noindent
this lagrangian give us the field equations
\begin{equation}
\label{1.2}
\epsilon^{\mu\nu\rho\sigma} \gamma_5 \gamma_\nu \partial_\rho \Psi_\sigma
= 0 \,.
\end{equation}
\noindent
It is however well known that the usual minimal electromagnetic
prescription for the Dirac field does not work adequatelly for this spin
$3/2$ field.
In fact if one couples minimally this field with electromagnetism, then several
physical inconsistencies arise of which the most remarkable is the appearance
of superluminal speed for the particles \cite{Velo}.

By demanding that the scattering amplitudes should have a good high energy
behaviour, Weinberg \cite{Weinberg} showed that the gyromagnetic ratio should
be $g = 2$ for arbitrary spin particles. Following a consistent procedure
for constructing the lagrangians for higher spin massive particles interacting
with the electromagnetic field, Ferrara {\it et al} \cite{Ferrara} obtained a
gyromagnetic ratio $g = 2$. As a result, their equations of motion contain an
extra dipole term that can be implemented at the tree level, thus modifying
the usual minimal electromagnetic coupling. A very important feature of this
extra dipole term is that, as shown by Ferrara {\it et al}, it avoids the
physical inconsistencies for spin $3/2$ particles described in
ref. \cite{Velo}.

On the other hand, recently two of the authors have constructed a theory of
the classical supersymmetric spin $3/2$ particle \cite{Nieto} in analogy with
the classical supersymmetric spin $1/2$ particle formalism developed by Galvao
and Teitelboim \cite{Galvao}. The usual $\gamma$ matrices were realized as
grassman variables and it was shown that the generalized Poisson bracket of
the Rarita-Schwinger constraint in flat space-time gives the full linearized
Einstein field equation.

If in the spin $3/2$ field differential operator we preserve the $\gamma$
matrices then the same result follows, but instead of the generalized Poisson
bracket, the anticommutator of the R-S operator must be used. This is
{\it not} a consequence of the well known result in canonical supergravity
\cite{Tei,Tab},
where it was shown that the supersymmetry constraint is the square root of
the usual constraint in canonical general relativity. This last
procedure involves only some of the dynamical equations, in contrast with the
relations found in ref. \cite{Nieto} which relates the complete set of
linearized Einstein field equations and the Rarita-Schwinger field equations.

The result of the paper mentioned above showed that the Rarita-Schwinger
equation is related with linearized gravity as the Dirac equation is related
to the Klein-Gordon equation.
Thus, following this analogy and knowing how gravity should couple with matter
one would expect to be able to find out the way matter couples with the spin
$3/2$ field in flat space-time.

The last point is the main guide for this work, because in principle we can
add a matter tensor on the right side of the linearized Einstein field
equations and investigate its ``square root'' in a similar manner to that
developed in \cite{Nieto}. As a result of this procedure a modified R-S
equation
will arise. Obviously, this ``square root'' must include the terms of
interaction with the gauge and matter fields, and these terms of interaction
will give us the information for the coupling of spin $3/2$ fields with any
kind of gauge fields and matter, particularly electromagnetism or Yang-Mills
fields.

This paper is organized as follows: in section 2 we generalize the four indices
differential operator representing the linearized general relativity equations
\cite{Nieto} in order to include electromagnetism and non-abelian Yang-Mills
fields.
Based on the particular form of this extra matter term in the linearized
Einstein field equations we search for the corresponding extra terms in the
Rarita-Schwinger equation which when squared will produce the implemented
matter term. As a result, we find the interaction for the spin $3/2$
field with electromagnetism and Yang-Mills fields. It is to be remarked that
this modified R-S equation when squared does not reproduce only the desired
extra term in the linearized gravity equations, but there appear extra terms.
This is not surprising because the relationship between the R-S equation with
interaction and the linearized Einstein field equations with matter is similar
to that existing between the Dirac equation with interaction and the
Klein-Gordon equation with a potential, where the $L S$ coupling term appears.
What we get is a kind of generalized ``hamiltonian'', for the linearized
gravity equations with a four indices generalized matter tensor.

Our result shows that the R-S equation is related in a direct manner with
linearized gravity. Formally, the latter is the ``hamiltonian'' and the
former one is its corresponding linear differential equation. On the other
hand our spin $3/2$ equation plays the role of some kind of supercharges. It is
then natural to think in Supergravity (SG), being the theory that incorporates
gravity, the spin $3/2$ field and gauge and matter fields.

We expect that by linearizing Supergravity we will be able to obtain the same
kind of results that in the case without matter \cite{Nieto}, and also when
interaction with matter is present. As mentioned, the relation we want to get
is between the full linearized R-S equations and linearized gravity and not
between the canonical constraints of supergravity \cite{Tei, Tab}. The free
case (without interaction) should correspond to linearized Supergravity N=1.
This is performed in section 3.

In the next two sections we also linearize Supergravity N=2 (section 4) and
N=4 (section 5). We show that the interaction found in section 2
for the electromagnetic field is similar to that obtained from supergravity
and is essentially also the same that Ferrara {\it et al} \cite{Ferrara}
found by a different procedure.
For the case of N=4 by imposing the scalar field equal to zero and neglecting
the spin $1/2$ field, we get the interaction term for a non-abelian Yang-Mills
field which has the same structure as that announced in section 2 .
However, in these cases there are correspondingly two and four spin
$3/2$ fields. The appearance of more spin $3/2$ fields is directly related
with the fact that in these last two cases we are treating with an enlarged
supersymmetry.

The R-S constraints resulting from supergravity are a very similar alternative
to that of taking the square root of linearized gravity in section 2.
However, they are very much more complicated to deal with.

Very probably, the existing relation between the full R-S and the linearized
Einstein field equations with matter had not been discovered from SG because
of its complicated structure when one includes matter and also because of its
necesary linearization. Given that gravity couples with all kinds of matter,
and being in our approach the spin $3/2$ field the square root of its
linearized limit, it is possible to extend our procedure in order to include
in the interaction with the R-S field any kind of gauge and matter fields,
obtaining in this way much more simple models than those obtained from
supergravity.

Following this prescription, it is in principle possible to construct
particular models considering interactions of phenomenological
interest including the mass of the spin $3/2$ field. This is the subject of
future work.


\section{Non-Minimal Coupling for Spin $3/2$ Fields \\
from Linearized Gravity with Matter}
\setcounter{equation}{0}
As mentioned in the introduction, we have on one hand the linearized Einstein
field equations  and on the other hand, we have the Rarita-Schwinger equations
as their square root. Thus it is natural to think that we can put an energy
momentum tensor for these Einstein field equations and obtain its square root
in order to investigate the possible coupling of the spin $3/2$ field with
gauge fields.

It has been shown in ref. \cite{Nieto} that if one associates to the R-S
equation the constraint
\begin{equation}
\label{2.1}
\hat{\cal S}^{\alpha\mu} \equiv \epsilon^{\alpha\mu\rho\sigma} \hat
\theta_\rho \hat P_\sigma = 0 \,,
\end{equation}
\noindent
where $\hat \theta_\rho = {1\over \sqrt 2} \gamma_5 \gamma_\rho $ and
$\hat P_\sigma = -i \partial_\sigma$. Then the square of this constraint
turns out to be
\begin{equation}
\label{2.2}
\hat{\cal H}^{\alpha\beta}_{\mu\nu} =
\epsilon^{\alpha~\rho\sigma}_{~\mu} \epsilon^{\beta~\lambda\gamma}_{~\nu}
\eta_{\rho\lambda} \hat P_\sigma \hat P_\gamma \,,
\end{equation}
\noindent
and this term is the ``hamiltonian'' operator that acts over $h_{\alpha\beta}$
in standard linearized gravity, {\it i. e.},
\begin{equation}
\label{2.3}
\hat{\cal H}^{\alpha\beta}_{\mu\nu} h_{\alpha\beta} = 0 \,.
\end{equation}
\noindent
We first note that in the hamiltonian ~(\ref{2.2}) the momenta $\hat P_\sigma$
appear quadratically. Now we want to add a potential term to
$ \hat {\cal H}^{\alpha\beta}_{\mu\nu}$, obviously, this must also be a four
indices tensor $ {\cal T}^{\alpha\beta}_{\mu\nu}$, it should also have units
of energy (same as $P_\sigma^2$) and it should be possible to take its
square root in terms of the fields characterizing the matter under
consideration.

In particular, for the electromagnetic case, there are two tensors at our
dispossal, $F_{\mu\nu}$ and its dual. We know that with them it is possible to
construct expressions with units of energy in order to get a potential term.
On the other hand, the mathematical structure of Eq. (\ref{2.2}) dictate us
to accompany the $F^2$ terms by two Levi-Civitta tensors, {\it i. e.},
\begin{equation}
\label{2.4}
{\cal T}^{\alpha\beta}_{\mu\nu} \sim
\epsilon^{\alpha~\rho\sigma}_{~\mu} \epsilon^{\beta~\lambda\gamma}_{~\nu}
\big( A F_{\rho\sigma}F_{\lambda\gamma} + B F_{\rho\sigma}
\tilde F_{\lambda\gamma} + C \tilde F_{\rho\sigma}
\tilde F_{\lambda\gamma} \big)\,,
\end{equation}
\noindent
where  $\tilde F_{\rho\sigma}$ is the usual dual of $ F_{\rho\sigma}$, A, B
and C are constants.
This expression fits the requeriments of being quadratic in the electromagnetic
field and its dual. Besides the mathematical structure of
(\ref{2.4}) is analogous to that of (\ref{2.2}), so the combined system of
expressions allow us to factorize the Levi-Civitta symbols in the computation
of the square root of the resulting generalized hamiltonian containing the
free plus the potential parts $\hat {\cal H}^{\alpha\beta}_{\mu\nu} +
{\cal T}^{\alpha\beta}_{\mu\nu}$.

The square root of this hamiltonian plus potential terms must give us
the R-S equations with the interaction term, and it is obvious that when we
square this modified R-S equation, it must give the full ``hamiltonian'' that
will result in the  hamiltonian (\ref{2.2}) plus the potential
${\cal T}^{\alpha\beta}_{\mu\nu}$, but also new
interaction terms as happens in the Dirac and Klein-Gordon equations.

Nevertheless, $\gamma$ matrices should appear in
the generalized R-S equation, analogous to the $\hat \theta$'s in Eq.
(\ref{2.1})
since we are computing the squares also by means of anticommutators, besides
we are applying these constraints to four spinor fields of four components.

\noindent
Then the desired interaction term in the $3/2$ field operator ought be of the
form
\begin{equation}
\label{2.5}
F^{\mu\nu} + \Gamma \tilde F^{\mu\nu} \,,
\end{equation}
\noindent
where $\Gamma$ is in general a product of Dirac matrices.
For convenience we will take this $\Gamma$ as $\gamma_5$ because of the
algebra we are working with \cite{Nieto} and the fact that when we square this
term, it will give us directly quadratic terms in the electromagnetic field
tensor and its dual.

A natural generalization of this result to the case of non-abelian Yang-Mills
would be the tensor
\begin{equation}
\label{2.6}
T^{\alpha\beta}_{\mu\nu} \sim
\epsilon_{~ \mu}^{\alpha ~ \rho\sigma} \epsilon_{~ \nu}^{\beta ~ \lambda\gamma}
\big( A F^a_{\rho\sigma}F^a_{\lambda\gamma} + B F^a_{\rho\sigma}
\tilde F^a_{\lambda\gamma} + C \tilde F^a_{\rho\sigma}
\tilde F^a_{\lambda\gamma} \big)\,,
\end{equation}
\noindent
where $F^a_{\rho\sigma}$ is the Yang-Mills field tensor. The corresponding
interaction term will also be of the form of (\ref{2.5}).

We are now in a position to generalize the R-S constraint (\ref{2.1})
by implementing the interaction (\ref{2.5}). For the electromagnetic case
it turns out to be
\begin{equation}
\label{2.7}
\epsilon^{\alpha\beta\mu\nu} \big[ \hat \theta_\mu \hat P_\nu +
{\it g_1} \{ \tilde F_{\mu\nu} + \gamma_5 F_{\mu\nu} \} \big] = 0 \,,
\end{equation}
\noindent
where we have factorized the Levi-Civitta symbol and introduced a coupling
constant ${\it g_1}$ which should be related to the gravitational constant
$\kappa$ since this accompanies the generalized energy momentum tensor in
linearized gravity.

As already mentioned, we expect that the square of expression (\ref{2.7})
should give us the generalized hamiltonian containing the free plus the
potential parts $\hat {\cal H}^{\alpha\beta}_{\mu\nu}
+ {\cal T}^{\alpha\beta}_{\mu\nu} $, besides it must give us an extra
interaction term $\hat {\cal I}^{\alpha\beta}_{\mu\nu} $.

The free hamiltonian is clearly the one given by (\ref{2.2}), the next two
terms are given by
\begin{equation}
\label{2.8}
{\cal T}^{\alpha\beta}_{\mu\nu} = 2 {\it g_1}^2
\epsilon^{\alpha~\rho\sigma}_{~\mu} \epsilon^{\beta~\lambda\gamma}_{~\nu}
\bigg( F_{\rho\sigma}F_{\lambda\gamma} + \tilde F_{\rho\sigma}
\tilde F_{\lambda\gamma}
+ {1\over \sqrt 2} \hat \theta_5 \big( F_{\rho\sigma} \tilde F_{\lambda\gamma}
+ \tilde F_{\rho\sigma} F_{\lambda\gamma} \big) \bigg) \,,
\end{equation}
\noindent
and
\begin{equation}
\label{2.9}
\hat {\cal I}^{\alpha\beta}_{\mu\nu} = {\it g_1} \bigg( \hat \theta_\rho
\big( -i \tilde F_{\lambda\gamma,\sigma} + F_{\lambda\gamma} \hat P_\sigma
- {\sqrt 2} i \hat \theta_5 F_{\lambda\gamma,\sigma} \big) +
\hat \theta_\lambda \big( -i \tilde F_{\rho\sigma,\gamma} + F_{\rho\sigma}
\hat P_\gamma
- {\sqrt 2} i \hat \theta_5 F_{\rho\sigma,\gamma} \big) \bigg) \,,
\end{equation}
\noindent
here we have made the identification $\hat \theta_5 \equiv {1\over\sqrt 2}
\gamma_5$.
Because of the appearance of the commutation relations between
$\hat P_\sigma$ and
$F_{\lambda\gamma}$ we identify this term as the analogous to the extra
term appearing in the Dirac equation when minimal electromagnetic
coupling is introduced. It give us the spin-orbit
interaction for the magnetic moment of the spin $1/2$ particles.
This analogy lead us to interesting physical interpretation and it is the
subject of future work.
Besides it is interesting to comment  that $ {\cal T}^{\alpha\beta}_{\mu\nu} $
may be understood as part of a total energy momentum that contains also
$ \hat {\cal I}^{\alpha\beta}_{\mu\nu}$.

When considering the Yang-MIlls case we will necesarily handle some internal
symmetries proper of the non-abelian field under consideration, namely
\begin{equation}
\label{2.10}
F^{\alpha\beta} = \alpha^k A^{\alpha\beta}_k \,,
\end{equation}
\noindent
where the $\alpha_k$ are the generators of the symmetry under consideration
and
\begin{equation}
\label{2.11}
A_{\rho\sigma}^k = \partial_\rho A^k_\sigma - \partial_\sigma A^k_\rho
+ e_A \epsilon^{ijk} A^i_\rho A^j_\sigma \,,
\end{equation}
\noindent
is the non-abelian Yang-Mills field. Then the R-S corresponding constraint
must have a similar structure to that of (\ref{2.7}), namely
\begin{equation}
\label{2.12}
\epsilon^{\alpha\beta\mu\nu} \big[ \hat \theta_\mu \hat P_\nu +
{\it g_2} \{ \tilde F_{\mu\nu} + \Gamma F_{\mu\nu} \} \big] = 0 \,,
\end{equation}
\noindent
where also ${\it g_2}$ must be related to the gravitational constant. The same
previous procedure is followed in order to find out the corresponding
generalized hamiltonian. In this case the potential and the interaction terms
turn out to be essentially the same than in the electromagnetic case, only
we have to replace ${\it g_1}$ by ${\it g_2}$ and the electromagnetic field
for the Yang-Mills field (\ref{2.10}), that is
\begin{equation}
\label{2.13}
{\cal T}^{\alpha\beta}_{\mu\nu} = 2 {\it g_2}^2
\epsilon^{\alpha~\rho\sigma}_{~\mu} \epsilon^{\beta~\lambda\gamma}_{~\nu}
\bigg( F_{\rho\sigma}F_{\lambda\gamma} + \tilde F_{\rho\sigma}
\tilde F_{\lambda\gamma}
+ {1\over \sqrt 2} \hat \theta_5 \big( F_{\rho\sigma} \tilde F_{\lambda\gamma}
+ \tilde F_{\rho\sigma} F_{\lambda\gamma} \big) \bigg) \,,
\end{equation}
\noindent
and
\begin{equation}
\label{2.14}
\hat {\cal I}^{\alpha\beta}_{\mu\nu} = {\it g_2} \bigg( \hat \theta_\rho
\big( -i \tilde F_{\lambda\gamma,\sigma} + F_{\lambda\gamma} \hat P_\sigma
- {\sqrt 2} i \hat \theta_5 F_{\lambda\gamma,\sigma} \big) +
\hat \theta_\lambda \big( -i \tilde F_{\rho\sigma,\gamma} + F_{\rho\sigma}
\hat P_\gamma
- {\sqrt 2} i \hat \theta_5 F_{\rho\sigma,\gamma} \big) \bigg) \,,
\end{equation}
\noindent
notice that the internal index $a$ of (\ref{2.6}) does not appear in these
expressions, since it is already taken into account in (\ref{2.10}).
Also the coupling terms in $\hat {\cal I}^{\alpha\beta}_{\mu\nu}$ are to be
investigated in future work.

In the next three sections we will show that linearized Supergravity N=1, 2
and 4 provide similar results for the free case, the electromagnetic one and
the non-abelian Yang-Mills field correspondingly. However, the enlarged
supersymmetries involved in the last two cases enforces the appearance of more
spin $3/2$ fields and the rising of much more complicated expressions than
those obtained in this section.


\section{Spin $3/2$ and Linearized Einstein Fields \\
from Supergravity  N=1}
\setcounter{equation}{0}
We begin by using the langrangian for {\it Supergravity N=1} given in ref.
\cite{PVN}
\begin{equation}
\label{3.1}
{\cal L} = -{ e \over 2}{\cal R} -
{ 1 \over 2}{\bar\Psi_\mu}\epsilon^{\mu\nu\rho\sigma}\gamma_5\gamma_\nu
D_\rho \Psi_\sigma\,,
\end{equation}
\noindent
where $e$ is the determinant of the tetrad, ${\cal R}$ is the curvature,
$\Psi_\mu$ is the gravitino field (spin $3/2$), and $D_\rho$ is the covariant
derivative including the spin connection.

We linearize the equations of motion by dropping quadratic and higher
terms since the spin connection in $D_\rho$ contains $\Psi^2$ and
$h_{\alpha\beta}$ that are applied to $\Psi_\mu$. After that we find once more
\begin{equation}
\label{3.2}
\epsilon^{\mu\nu\rho\sigma} \gamma_5 \gamma_\nu \partial_\rho \Psi_\sigma = 0
\,,
\end{equation}
\noindent
we can associate to Eq. (\ref{3.2}) the operator
\begin{equation}
\label{3.3}
\hat{\cal S}^{\mu\nu} \equiv \epsilon^{\mu\nu\rho\sigma} \hat
\theta_\rho \hat P_\sigma = 0 \,,
\end{equation}
\noindent
where $\hat \theta_\rho $ and $\hat P_\sigma $ are given in the preceding
section.

Then, by using the algebra of the anticommutators of $\hat \theta_\mu$
\cite{Nieto} given by
\begin{equation}
\label{3.4}
\{ \hat \theta_\mu, \hat \theta_\nu \} = \eta_{\mu\nu} \,,
\end{equation}
\noindent
we obtain
\begin{equation}
\label{3.5}
\{ \hat{\cal S}^a_\mu, \hat{\cal S}^\beta_\nu \}   =
\hat{\cal H}^{\alpha\beta}_{\mu\nu}\,,
\end{equation}
\noindent
where
\begin{equation}
\label{3.6}
\hat{\cal H}^{\alpha\beta}_{\mu\nu} =
\epsilon^{\alpha~\rho\sigma}_{~\mu} \epsilon^{\beta~\lambda\gamma}_{~\nu}
\eta_{\rho\lambda} \hat P_\sigma \hat P_\gamma \,,
\end{equation}
\noindent
is the ``hamiltonian'' operator that acts over $h_{\alpha\beta}$ in standard
linearized gravity as described in the preceding section and in ref.
\cite{Nieto}.

As claimed before, the Rarita-Schwinger equations in flat space-time turn out
be the square root of the linearized Einstein field equations. Obviously, we
have no contribution of any matter field.


\section{Electromagnetic Interaction for Spin $3/2$ Fields \\
from Supergravity N=2}
\setcounter{equation}{0}
We present the electromagnetic coupling of spin $3/2$ fields by using
{\it Supergravity N=2} \cite{PVN}, since it naturally
incorporates the graviton, the electromagnetic field and two gravitinos.

The lagrangian for Supergravity N=2 is
\begin{eqnarray}
\label{4.1}
{\cal L}  & = &  -{ e \over 2}{\cal R} -
{ 1\over 2}{\bar\Psi^i_\mu}\epsilon^{\mu\nu\rho\sigma}\gamma_5 \gamma_\nu
D_\rho{\Psi^i_\sigma}
- { e \over 4}F_{\alpha\beta} F^{\alpha\beta} \\
& + & {\kappa \over 4\sqrt2} \bar\Psi^i_\mu \big[e(F^{\mu\nu} +
\hat F^{\mu\nu}) + \gamma_5({\tilde F^{\mu\nu}} +
{\skew4 \tilde {\hat F}^{\mu\nu}})\big]
{\Psi^j_\nu}\epsilon^{ij}, \nonumber \\
\nonumber
\end{eqnarray}
\noindent
where the elements of this lagrangian are the same that in the precceding
section and $\tilde F^{\mu\nu} ={\scriptstyle{1\over 2}}
\epsilon^{\mu\nu\alpha\beta} F_{\alpha\beta}$
is the dual of the usual electromagnetic fields strenght tensor and
\\
$ \hat F_{\mu\nu} = \partial_\mu A_\nu - \partial_\nu A_\mu -
{\kappa \over 2\sqrt2}[\Psi^i_\mu \Psi^j_\nu - \Psi^i_\nu \Psi^j_\mu]
\epsilon^{ij}$ is the supercovariant curl.
The appearance of the $ \epsilon^{ij}$ in the lagrangian and in the above
equation has to do with the fact that, as mentioned in the introduction, we
are leading with a larger supersymmetry. This is reflected in the above
expression, in which the $ \epsilon^{ij}$ mix up the interaction of the two
gravitinos.

After linearizing the resulting equations of motion following the same
proccedure of the previous section, they are reduced to
\begin{equation}
\label{4.2}
\epsilon^{\alpha\mu\nu\beta}\gamma_5\gamma_\mu \partial_\nu \Psi^i_\beta
- {\kappa \epsilon^{ij} \over {\sqrt 2}} \big[
F^{\alpha\beta} + {\scriptstyle{1\over 2}}\gamma_5
\epsilon^{\alpha\beta\rho\sigma} F_{\rho\sigma} \big]
\Psi^j_\beta  = 0 \,.
\end{equation}
\noindent
Notice that the term in squared brackets $F^{+\alpha\beta}
\equiv F^{\alpha\beta} + {\scriptstyle{1\over 2}}
\gamma_5 \epsilon^{\alpha\beta\rho\sigma} F_{\rho\sigma} $
is exactly the interaction previously found in section 2. This term has been
also deduced from a consistent gauge field theory and is precisely the dipole
term found by Ferrara {\it et al} \cite{Ferrara} following a lagrangian
procedure based on demanding $g=2$ for arbitrary spin paricles. In their
article they have shown that this term cancells divergences and avoids
superluminal velocities in systems of spin $3/2$ particles.
For these reasons the solutions associated to (\ref{4.2}) and the previously
proposed interaction (\ref{2.7}) should not have any physical inconsistencies.

Equation (\ref{4.2}) can be shown to be the generalized Rarita-Schwinger
equation
\begin{equation}
\label{4.3}
\epsilon^{\alpha\beta\mu\nu} \big[ \delta_{ij} \hat \theta_\mu \hat P_\nu +
{\it g_1} \epsilon_{ij} \big( \tilde F_{\mu\nu} +
{\sqrt 2} \hat \theta_5 \tilde F_{\mu\nu} \big) \big] \Psi^j_\beta = 0 \,,
\end{equation}

\noindent
where ${\it g_1} = {i \kappa \over 8} $ is the coupling constant, and as
mentioned in section 2, it is related to the gravitational constant $\kappa$.

Thus, we can associate to Eq. (\ref{4.3}) the constraint
\begin{equation}
\label{4.4}
\hat{\cal S}^{\alpha\beta}_{ij} = \epsilon^{\alpha\beta\mu\nu}
\big[ \delta_{ij} \hat \theta_\mu \hat P_\nu +
{\it g_1} \epsilon_{ij} \big( \tilde F_{\mu\nu} +
{\sqrt 2} \hat \theta_5 \tilde F_{\mu\nu} \big) \big] = 0 \,,
\end {equation}
\noindent
which is essentially the same constraint deduced in (\ref{2.7}) but here
$\delta$'s and $\epsilon$'s factors appear because of the enlarged
supersymmetry we are working with.

By using the anticommutators of section 3 and
\begin{equation}
\label{4.5}
\{ \hat \theta_5,  \hat \theta_5 \} = 1 \,,
\end{equation}
\noindent
we get the algebra
\begin{eqnarray}
\label{4.6}
\big\{ \hat {\cal S}^{\alpha ~ij}_{~\mu} , \hat {\cal S}^{\beta ~kl}_{~\nu}
\big\} & = &
\epsilon^{\alpha~\rho\sigma}_{~\mu} \epsilon^{\beta~\lambda\gamma}_{~\nu}
\bigg[ \eta_{\rho\lambda} \hat P_\sigma \hat P_\gamma \delta^{ij} \delta^{kl}
\nonumber \\
& + & 2 {\it g_1}^2 \bigg( F_{\rho\sigma}F_{\lambda\gamma}
+ {1 \over \sqrt 2} \hat \theta_5 \big( F_{\rho\sigma}\tilde F_{\lambda\gamma}
+  F_{\lambda\gamma}\tilde F_{\rho\sigma} \big)
+  \tilde F_{\rho\sigma} \tilde F_{\lambda\gamma} \bigg)
\epsilon^{ij} \epsilon^{lk}  \nonumber \\
& + & {\it g_1} \bigg(
\delta^{ij} \epsilon^{lk} \hat \theta_\rho \bigg(
- i \tilde F_{\lambda\gamma, \sigma}
+ 2 \tilde F_{\lambda\gamma} \hat P_\sigma  - {\sqrt 2} i \hat \theta_5
F_{\lambda\gamma, \sigma} \bigg) \\
& + & \epsilon^{ij} \delta^{lk} \hat \theta_\lambda
\bigg( -i \tilde F_{\rho\sigma, \gamma} + 2 \tilde F_{\rho\sigma}
\hat P_\gamma
- {\sqrt 2} i \hat \theta_5 F_{\rho\sigma, \gamma} \bigg)
\bigg) \bigg]. \nonumber \\
\nonumber
\end{eqnarray}
The first term in the last equation is the hamiltonian for linearized gravity
discussed before up to delta factors

\begin{equation}
\label{4.7}
\hat {\cal H}^{\alpha\beta}_{\mu\nu} \delta^{ij} \delta^{lk} =
\epsilon^{\alpha~\rho\sigma}_{~\mu} \epsilon^{\beta~\lambda\gamma}_{~\nu}
\eta_{\rho\lambda} \hat P_\sigma \hat P_\gamma \delta^{ij} \delta^{lk} \,.
\end{equation}

\noindent
The second of these terms is precisely the generalized energy momentum for the
electromagnetic field found in section 2 up to epsilon factors, that is

\begin{equation}
\label{4.8}
{\cal T}^{\alpha\beta}_{\mu\nu} \epsilon^{ij} \epsilon^{kl} =
2 {\it g_1}^2 \bigg( F_{\rho\sigma}F_{\lambda\gamma}
+ {1 \over \sqrt 2} \hat \theta_5 \big( F_{\rho\sigma}\tilde F_{\lambda\gamma}
+  F_{\lambda\gamma}\tilde F_{\rho\sigma} \big)
+  \tilde F_{\rho\sigma} \tilde F_{\lambda\gamma} \bigg)
\epsilon^{ij} \epsilon^{lk} \,.
\end{equation}

\noindent
This result contributes to corroborate the proposed generalized energy
momentum tensors (\ref{2.4}) and (\ref{2.8}).
However, there is an extra term of interaction, this term is
\begin{eqnarray}
\label{4.9}
\hat {\cal I}^{\alpha\beta ijkl}_{\mu\nu}  & \equiv & {\it g_1}
\epsilon^{\alpha~\rho\sigma}_{~\mu} \epsilon^{\beta~\lambda\gamma}_{~\nu}
\bigg(
\delta^{ij} \epsilon^{lk} \hat \theta_\rho \bigg(
- i \tilde F_{\lambda\gamma, \sigma}
+ 2 \tilde F_{\lambda\gamma} \hat P_\sigma  - {\sqrt 2} i \hat \theta_5
F_{\lambda\gamma, \sigma} \bigg) \nonumber \\
& + & \epsilon^{ij} \delta^{lk} \hat \theta_\lambda
\bigg( -i \tilde F_{\rho\sigma, \gamma} + 2 \tilde F_{\rho\sigma}
\hat P_\gamma
- {\sqrt 2} i \hat \theta_5 F_{\rho\sigma, \gamma} \bigg)
\bigg), \\
\nonumber
\end{eqnarray}
\noindent
as already mentioned in section 2 this expression contains the information of
terms like the spin coupling and it deserves further study.


\section{Yang-Mills Interaction for Spin $3/2$ Fields \\
from Supergravity N=4}
\setcounter{equation}{0}
Now we are in a position to explore the coupling of a Yang-Mills field
with Rarita-Schwinger field. As a Supergravity model that
involves a non-abelian Yang-Mills field we take {\it Supergravity N=4}
\cite{PVN, Freedman}.

In the philosophy of the preceding calculations we can asociate a
constraint to the field equations for the gravitinos, after eliminating
the scalar and the spin $1/2$ fields, it turns out to be
\begin{equation}
\label{5.1}
\hat {\cal S}^{\alpha\beta}_{ij} \equiv \epsilon^{\alpha\beta}_{~~\mu\nu}
\bigg( \delta_{ij}\hat \theta^\mu \hat P^\nu
+  {\it g_2} \big( \tilde F^{\mu\nu}_{(ij)}
- {\sqrt 2} i \hat \theta_5 F^{\mu\nu}_{(ij)} \big) \bigg)
+ { i \delta_{ij} \over 2 \kappa} \big( e_A
+ {\sqrt 2} i e_B \hat \theta_5 \big) \sigma^{\alpha\beta} \,,
\end{equation}
\noindent
where now the coupling constant ${ i \kappa \over 4} $ and the
contribution of the non-abelian field is given by
\begin{equation}
\label{5.2}
{\cal F}^{\rho\sigma}_{(ij)} = \alpha^k_{(ij)} A^{\rho\sigma}_k + {\sqrt 2} i
\hat \theta_5  \beta^k_{(ij)} B^{\rho\sigma}_k\,,
\end{equation}
\noindent
and
\begin{eqnarray}
A_{\rho\sigma}^k & = & \partial_\rho A^k_\sigma - \partial_\sigma A^k_\rho
+ e_A \epsilon^{ijk} A^i_\rho A^j_\sigma, \\
B_{\rho\sigma}^k & = & \partial_\rho B^k_\sigma - \partial_\sigma B^k_\rho
+ e_B \epsilon^{ijk} B^i_\rho B^j_\sigma, \\
\nonumber
\end{eqnarray}
\noindent
are the non-abelian Yang-Mills fields. In Eqs. (\ref{5.1}) and (\ref{5.2})
the indices {\it i} and {\it j} refer to the four gravitinos and the
components of the matrices alpha and beta that generate the $SU(2)\times
SU(2)$ symmetry of supergravity N=4.
Notice that in the constraint (\ref{5.1}) we have preserved the notation
of ref. \cite{Freedman} and the difference in the definition of $\gamma_5$
is obvious with respect to that of the preceding sections. In order to recover
the notation used before we have to perform the sustitution
$-i \hat \theta_5 \rightarrow \hat \theta_5$, thus the constraint (\ref{5.1})
becomes
\begin{equation}
\label{5.5}
\hat {\cal S}^{\alpha\beta}_{ij} \equiv \epsilon^{\alpha\beta}_{~~\mu\nu}
\bigg( \delta_{ij}\hat \theta^\mu \hat P^\nu
+  {\it g_2} \big( \tilde F^{\mu\nu}_{(ij)}
+ {\sqrt 2} \hat \theta_5 F^{\mu\nu}_{(ij)} \big) \bigg)
+ { i \delta_{ij} \over 2 \kappa} \big( e_A
- {\sqrt 2} e_B \hat \theta_5 \big) \sigma^{\alpha\beta} \,,
\end{equation}
\noindent
By using anticommutators as in the preceding sections, we get the algebra
\begin{eqnarray}
\label{5.6}
\big\{ \hat {\cal S}^\alpha_{~\mu ij} , \hat {\cal S}^\beta_{~\nu kl} \big\}
& = &
\epsilon^\alpha_{~\mu\rho\sigma} \epsilon^\beta_{~\nu\lambda\gamma}
\bigg[ \eta^{\rho\lambda} \hat P^\sigma \hat P^\gamma \delta_{ij} \delta_{kl}
\nonumber \\
& + & 2{\it g_2}^2 \bigg( F^{\rho\sigma}_{(ij)} F^{\lambda\gamma}_{(kl)}
+ {1\over \sqrt 2} \hat \theta_5 ( \tilde F^{\rho\sigma}_{(ij)}
F^{\lambda\gamma}_{(kl)}
+ F^{\rho\sigma}_{(ij)} \tilde F^{\lambda\gamma}_{(kl)})
+ \tilde F^{\rho\sigma}_{(ij)} \tilde F^{\lambda\gamma}_{(kl)} \bigg)
\nonumber \\
& + & {\it g_2} \bigg(
\delta_{ij} \hat \theta^\rho \big( - i  \tilde F^{\lambda\gamma,\sigma}_{(kl)}
+ 2 \tilde F^{\lambda\gamma}_{(kl)} \hat P^\sigma
- {\sqrt 2}i \hat \theta_5 F^{\lambda\gamma,\sigma}_{(kl)} \big) \nonumber \\
& + & ~~~\delta_{lk} \hat \theta^\lambda \big( - i
\tilde F^{\rho\sigma,\gamma}_{(ij)}
+ 2 \tilde F^{\rho\sigma}_{(ij)} \hat P^\gamma
- {\sqrt 2}i \hat \theta_5 F^{\rho\sigma,\gamma}_{(ij)} \big) \bigg) \bigg]
\nonumber
\nonumber \\
& + & {i \delta_{kl} \over 2 \kappa} \epsilon^\alpha_{~\mu\rho\sigma} \bigg[
\delta_{ij} \big(
  {\sqrt 2} \epsilon^{\rho\beta}_{~~\nu\tau} \hat \theta_5 \hat \theta^\tau
\hat P^\sigma
- {\sqrt 2} e_B \hat \theta_5 (\delta^\rho_{~\nu} \hat \theta^\beta -
\delta^{\rho\beta} \hat \theta_\nu) \hat P^\sigma \big)
\nonumber \\
& + & 2 {\it g_2} \big( e_A \tilde F^{\rho\sigma}_{(ij)} \sigma^\beta_{~\nu}
- {\sqrt 2} e_B \hat \theta_5 \tilde F^{\rho\sigma}_{(ij)} \sigma^\beta_{~\nu}
- e_B F^{\rho\sigma}_{(ij)} \sigma^\beta_{~\nu} \big) \bigg] \\
& + & {i \delta_{ij} \over 2 \kappa} \epsilon^\beta_{~\nu\lambda\gamma} \bigg[
\delta_{kl} \big(
  {\sqrt 2} \epsilon^{\lambda\alpha}_{~~\mu\tau} \hat \theta_5
\hat \theta^\tau \hat P^\gamma
- {\sqrt 2} e_B \hat \theta_5 (\delta^\lambda_{~\mu} \hat \theta^\alpha -
\delta^{\lambda\alpha} \hat \theta_\mu) \hat P^\gamma \big)
\nonumber \\
& + & 2 {\it g_2} \big( e_A \tilde F^{\lambda\gamma}_{(kl)}
\sigma^\alpha_{~\mu}
- {\sqrt 2} e_B \hat \theta_5 \tilde F^{\lambda\gamma}_{(kl)}
\sigma^\alpha_{~\mu}
- e_B F^{\lambda\gamma}_{(kl)} \sigma^\alpha_{~\mu} \big) \bigg]
\nonumber \\
& - & {\delta_{ij} \delta_{kl} \over 4 \kappa^2} \bigg[ \big( e_A^2 + e_B^2
\big)
\big( \eta^{\alpha\beta} \sigma_{\mu\nu} + \eta_{\mu\nu} \sigma^{\alpha\beta}
+ \delta^\alpha_{~\nu} \sigma^\beta_{~\mu} + \delta^\beta_{~\mu}
\sigma^{~\alpha}_{\nu} \big)
+ 2 e_B^2 \sigma^\alpha_{~\mu} \sigma^\beta_{~\nu}  \bigg].
\nonumber
\end{eqnarray}
\noindent
We can identify terms and the first of them is essentially the same
tensor found in (\ref{2.2}), it gives us the linearized operator for the
Einstein field equations up to delta factors. That is
\begin{equation}
\label{5.7}
\hat {\cal H}^{\alpha\beta}_{\mu\nu} \delta_{ij} \delta_{lk} =
\epsilon^{\alpha~\rho\sigma}_{~\mu} \epsilon^{\beta~\lambda\gamma}_{~\nu}
\eta_{\rho\lambda} \hat P_\sigma \hat P_\gamma \delta_{ij} \delta_{lk} \,.
\end{equation}
The second term is the Yang-Mills field energy momentum tensor obtained in
section 2, corresponding to the one deduced in (\ref{2.13}) by taking the
square root of linearized gravity. This term is:
\begin{eqnarray}
\label{5.8}
{\cal T}^{\alpha\beta}_{\mu\nu(ijkl)}  & = & 2 {\it g_2}^2
\epsilon^{\alpha~\rho\sigma}_{~\mu} \epsilon^{\beta~\lambda\gamma}_{~\nu}
\bigg( F_{\rho\sigma(ij)} F_{\lambda\gamma(kl)} \\
& + &  {1\over \sqrt 2} \hat \theta_5 ( \tilde F_{\rho\sigma(ij)}
F_{\lambda\gamma(kl)}
+ F_{\rho\sigma(ij)} \tilde F_{\lambda\gamma(kl)} )
+ \tilde F_{\rho\sigma(ij)} \tilde F_{\lambda\gamma(kl)}
\bigg), \nonumber \\
\nonumber
\end{eqnarray}
\noindent
in this case the latin indices refer to the components  of the
$\alpha$ and $\beta$ matrices mentioned before, these indices apply over the
different gravitinos.
\noindent
The following term is the corresponding to the interaction term (2.14) for the
Yang-Mills field in comparison to that obtained in section 2
(\ref{4.9})
\begin{eqnarray}
\label{5.9}
\hat {\cal I}^{\alpha\beta}_{\mu\nu(ijkl)}  & \equiv &
\epsilon^\alpha_{~\mu\rho\sigma} \epsilon^\beta_{~\nu\lambda\gamma}
\bigg[
{\it g_2} \bigg(
\delta_{ij} \hat \theta^\rho \big( - i  \tilde F^{\lambda\gamma,\sigma}_{(kl)}
+ 2 \tilde F^{\lambda\gamma}_{(kl)} \hat P^\sigma
- {\sqrt 2}i \hat \theta_5 F^{\lambda\gamma,\sigma}_{(kl)} \big) \nonumber \\
& + & ~~~\delta_{lk} \hat \theta^\lambda \big( - i
\tilde F^{\rho\sigma,\gamma}_{(ij)}
+ 2 \tilde F^{\rho\sigma}_{(ij)} \hat P^\gamma
- {\sqrt 2}i \hat \theta_5 F^{\rho\sigma,\gamma}_{(ij)} \big) \bigg) \bigg]
\nonumber \\
& + & {i \delta_{kl} \over 2 \kappa} \epsilon^\alpha_{~\mu\rho\sigma} \bigg[
\delta_{ij} \big(
  {\sqrt 2} \epsilon^{\rho\beta}_{~~\nu\tau} \hat \theta_5 \hat \theta^\tau
\hat P^\sigma
- {\sqrt 2} e_B \hat \theta_5 (\delta^\rho_{~\nu} \hat \theta^\beta -
\delta^{\rho\beta} \hat \theta_\nu) \hat P^\sigma \big)
\nonumber \\
& + & 2 {\it g_2} \big( e_A \tilde F^{\rho\sigma}_{(ij)} \sigma^\beta_{~\nu}
- {\sqrt 2} e_B \hat \theta_5 \tilde F^{\rho\sigma}_{(ij)} \sigma^\beta_{~\nu}
- e_B F^{\rho\sigma}_{(ij)} \sigma^\beta_{~\nu} \big) \bigg] \\
& + & {i \delta_{ij} \over 2 \kappa} \epsilon^\beta_{~\nu\lambda\gamma} \bigg[
\delta_{kl} \big(
 {\sqrt 2} \epsilon^{\lambda\alpha}_{~~\mu\tau} \hat \theta_5 \hat \theta^\tau
\hat P^\gamma
- {\sqrt 2} e_B \hat \theta_5 (\delta^\lambda_{~\mu} \hat \theta^\alpha -
\delta^{\lambda\alpha} \hat \theta_\mu) \hat P^\gamma \big)
\nonumber \\
& + & 2 {\it g_2} \big( e_A \tilde F^{\lambda\gamma}_{(kl)}
\sigma^\alpha_{~\mu}
- {\sqrt 2} e_B \hat \theta_5 \tilde F^{\lambda\gamma}_{(kl)}
\sigma^\alpha_{~\mu}
- e_B F^{\lambda\gamma}_{(kl)} \sigma^\alpha_{~\mu} \big) \bigg]
\nonumber \\
& - & {\delta_{ij} \delta_{kl} \over 4 \kappa^2} \bigg[
\big( e_A^2 + e_B^2 \big)
\big( \eta^{\alpha\beta} \sigma_{\mu\nu} + \eta_{\mu\nu} \sigma^{\alpha\beta}
+ \delta^\alpha_{~\nu} \sigma^\beta_{~\mu} + \delta^\beta_{~\mu}
\sigma^{~\alpha}_{\nu} \big)
+ 2 e_B^2 \sigma^\alpha_{~\mu} \sigma^\beta_{~\nu}  \bigg]. \nonumber \\
\nonumber
\end{eqnarray}
\noindent
Obviously these interaction terms are much more complicated than those obtained
in (\ref{2.14}). They provide however a different approach to our square root
procedure developed in section 2, and they should be the subject of future
research.


\section{Conclusions}
\setcounter{equation}{0}
We have discussed the problem of the non-minimal electromagnetic and Yang-Mills
coupling for the spin $3/2$ field. We took advantage of a precceding work
where we showed
the fact that the Rarita-Schwinger equations in flat space-time are the square
root of the linearized Einstein field equations. This fact was used in order to
find out the way gauge fields should couple with spin $3/2$ fields by
implementing
appropriate generalized energy momentum tensors on the linearized Einstein
field equations.
In particular for the electromagnetic and Yang-Mills fields, these generalized
energy momentum tensors were build up and inserted in the linear gravity
equations. This procedure allowed us to take the square root of the system
consisting of linearized gravity plus matter, thus obtaining modified
Rarita-Schwinger equations containing the pursued coupling with gauge fields.
Similar R-S equations were obtained by linearizing Supergravity which
enforces our proposal. When we square these equations, linearized gravity
plus the generalized energy momentum arise, but also expected interaction
terms analogous to the spin-orbit coupling terms of the Dirac and
Klein-Gordon equations. These interesting terms deserve future study.

In the case of electromagnetic interaction, the coupling obtained by us
coincides with the one found by Ferrara {\it et al} \cite{Ferrara}  by a very
different procedure, namely by demanding $g = 2$ at the tree level for
arbitrary spin particles. They obtained an extra dipole term in the equations
for these fields.
This dipole term avoids the bad energy behavior in systems of spin $3/2$
particles and the physical inconsistencies discussed in ref. \cite{Velo}.

The electromagnetic and non-abelian Yang-Milss fiedls interaction terms
are both of the form
\begin{equation}
\label{6.1}
Field + \gamma_5 Dual~ Field\,.
\nonumber
\end{equation}
\noindent
In our square root approach it is always possible to add to the above
interaction any other gauge and matter fields. It is the subject of future
work to propose particular interactions of possible phenomenological
implications.

It is interesting to mention that a term of similar structure was implemented
by Cucchieri, Porrati and Deser \cite{Cucc} by analyzing the gravitational
coupling for higher spin massive fields, where the {\it Field} of the above
expression (\ref{6.1})is the Riemman tensor itself. This happens, in
particular, when spin $5/2$ is considered.

\vfill\eject
\section*{Acknowledgements}

This work was supported in part by a CONACyT grant 4862 E9406, a Catedra
Patrimonial de Excelencia Nivel II 940055 and by Coordinaci\'on de la
Investigaci\'on Cient\'{\i}fica de la UMSNH. VMV is supported by a CONACyT
and a SNI Graduate Studentship. We thank J. L. Lucio for helpful discussions.

\frenchspacing



\begin{thebibliography}{99}
\bibitem{RS} W. Rarita and J. Schwinger {\bf Phys. Rev. 60}  61 (1941)
\bibitem{LL} J. Leite Lopes, D. Spheeler and N. Fleury {\bf Lett. Nuovo
Cimento 35} 60 (1982)
\bibitem{Velo}G. Velo and D. Zwanziger {\bf Phys. Rev 186} 25 (1969)
\bibitem{Weinberg}S. Weinberg, in {\it Lectures on Elementary Particles and
Quantum Field Theory} Proceedings of the Summer Institute, Brandeis
University, 1970, edited by S. Deser (MIT Press, Cambridge, MA, 1970), Vol. I.
\bibitem{Ferrara}S. Ferrara, M. Porrati and V. L. Telegdi {\bf Phys. Rev. D 46}
3529 (1992)
\bibitem{Nieto}J. A. Nieto and O. Obreg\'on {\bf Phys. Lett. A 175} 11 (1993)
\bibitem{Galvao}C. Galvao and C. Teitelboim {\bf J. Math. Phys. 21} 1863 (1980)
\bibitem{Tei}C. Teitelboim {\bf Phys. Rev. Lett. 38} 1106 (1977)
\bibitem{Tab}R. Tabenski and C. Teitelboim {\bf Phys. Lett. 69B} 453 (1977)
\bibitem{PVN}P. van Nieuwenhuizen {\bf Physics Reports 68} No. 4 (1981)
189-398 North-Holland Publishing Company.
\bibitem{Freedman}D. Z. Freedman and J. H. Schwarz {\bf Nucl. Phys. B137}
333 (1978)
\bibitem{Cucc}A. Cucchieri, M. Porrati and S. Deser {\bf Phys. Rev. D 51}
4543 (1995)

\end{thebibliography}
\end{document}